\documentclass[aps,prc,twocolumn,superscriptaddress]{revtex4-2}

\usepackage{graphicx}
\usepackage{dcolumn}
\usepackage{bm}
\usepackage{subfigure}
\usepackage{mathrsfs}
\usepackage{multirow}
\usepackage{amsmath}

\usepackage[colorlinks,
linkcolor={red!60!black!},
anchorcolor={green!80!black!},
urlcolor={blue!80!black!},
citecolor={blue!50!black!}]{hyperref}
\usepackage[table]{xcolor}

\gdef\N3LOlnl{N$^3$LO+3N$_{\rm lnl}$}

\gdef\BE2{$B({\rm E}2)$}

\begin{document}

\title{Ab initio computations of strongly deformed nuclei around $^{80}$Zr}

\author{B. S. Hu} 
\affiliation{National Center for Computational Sciences, Oak Ridge National Laboratory, Oak Ridge, Tennessee 37831, USA}
\affiliation{Physics Division, Oak Ridge National Laboratory, Oak Ridge, Tennessee 37831, USA}

\author{Z. H. Sun} 
\affiliation{Physics Division, Oak Ridge National Laboratory, Oak Ridge, Tennessee 37831, USA} 

\author{G. Hagen} 
\affiliation{Physics Division, Oak Ridge National Laboratory, Oak Ridge, Tennessee 37831, USA} 
\affiliation{Department of Physics and Astronomy, University of Tennessee, Knoxville, Tennessee 37996, USA}

\author{T. Papenbrock} 
\affiliation{Department of Physics and Astronomy, University of Tennessee, Knoxville, Tennessee 37996, USA} 
\affiliation{Physics Division, Oak Ridge National Laboratory, Oak Ridge, Tennessee 37831, USA}


\begin{abstract}
Nuclei around $N\approx Z\approx 40$ are strongly deformed and exhibit coexistence of shapes. These phenomena have challenged nuclear models. 
Here we perform ab initio coupled-cluster computations of low-lying collective states and electromagnetic quadrupole transitions of the even-even nuclei $^{72}$Kr, $^{76,78}$Sr, $^{78,80}$Zr and $^{84}$Mo starting from chiral nucleon-nucleon and three-nucleon forces. Our calculations reproduce the coexistence of oblate and prolate shapes in these nuclei, yield rotational bands and strong electromagnetic transitions, but are not accurate for some observables and nuclei. 
These results highlight the advances and challenges of ab initio computations of heavy deformed nuclei. 

\end{abstract}

\maketitle

{\it Introduction.$-$}
Atomic nuclei in the vicinity of the neutron deficient nucleus $^{80}$Zr have a rich and interesting structure; they are often strongly deformed~\cite{lister1987,llewellyn2020}, exhibit shape coexistence and mixing~\cite{bouchez2003,heyde2011,wimmer2020,garrett2022}, and pose a challenge to nuclear models~\cite{hamilton1984,nazarewicz1985,petrovici1996,reinhard1999,langanke2003b,sun2004,rodriguez2011,zheng2014,miyahara2018,kaneko2021,thakur2021,chimanski2023}. 
The energy ratio $R_{4/2} \equiv E(4_1^+)/E(2_1^+)$ and reduced electromagnetic quadrupole transition probability \BE2~$\equiv B({\rm E}2;2^+_1 \rightarrow 0^+_1)$ highlight the emergence of deformation around $N\approx Z\approx40$~\cite{lister1982,gade2005,hasegawa2007,heyde2011,lemasson2012,llewellyn2020}. In particular, there is a transition from oblate ground state shapes in $^{68}$Se and $^{72}$Kr \cite{bouchez2003,gade2005,obertelli2009,iwasaki2014} to prolate ones in $^{76}$Sr and $^{80}$Zr \cite{hamilton1984,nazarewicz1985,lister1987,reinhard1999,nacher2004,lemasson2012,llewellyn2020,kaneko2021}. These nuclei also exhibit coexisting shapes. Beyond mean-field calculations, for example, predicted that $^{80}$Zr has five nearly degenerate nuclear shapes \cite{rodriguez2011}. In $^{72}$Kr, there is mixing between oblate and prolate shapes within the ground-state rotational band~\cite{deangelis1997,wimmer2020}. 

Thus, these nuclei provide us with unique phenomena to test nuclear models and methods. 
So far, shape coexistence and deformation in the heavy $N\approx Z$ region have been studied using mean-field methods \cite{nazarewicz1985,reinhard1999,lemasson2012,llewellyn2020,rodriguez2011,hamaker2021} and large-scale shell model computations \cite{langanke2003b,sun2004,hasegawa2007,zuker2015,kaneko2021}. Although these calculations have guided and interpreted experiments, the data clearly challenge the theory. In $^{80}$Zr, for instance, the mean-field computations ~\cite{delaroche2010,rodriguez2011} overpredict the \BE2~\cite{llewellyn2020}, and several calculations yield a spherical (and not a deformed) ground state~\cite{miyahara2018,thakur2021,chimanski2023}. The recent high-precision mass measurement of this nucleus revealed a large deformed shell gap which challenged nuclear models~\cite{hamaker2021}. This makes it interesting to see how ab initio methods fare in this region of the nuclear chart.  

Ab initio computations have advanced tremendously in recent years, see Refs.~\cite{morris2018,gysbers2019,yao2020,stroberg2021,hu2022} for examples and Ref.~\cite{hergert2020} for a recent review. In this Letter, we compute nuclei around $^{80}$Zr ab initio, following the interpretation of this expression by \textcite{Ekstrom:2022yea}. We study the structure and electric quadrupole transitions of the nuclei $^{72}$Kr, $^{76,78}$Sr, $^{78,80}$Zr and $^{84}$Mo based on chiral nucleon-nucleon and three-nucleon forces.  

Our calculations start from axially deformed Hartree-Fock states, and we calculate the energy using single reference coupled cluster theory~\cite{bartlett2007,novario2020,novario2023}. The broken rotational symmetry is then restored through angular momentum projection~\cite{qiu2017,qiu2018,qiu2019,hagen2022}. This approach captures both short- and long-range correlations \cite{sun2024}. As we shall see, the ab initio results are competitive with those from axially symmetric (beyond) mean-field methods and -- like those -- are also challenged by the data. The calculations of this paper target nuclei that are about twice as heavy as the deformed neon and magnesium nuclei in the first island of inversion~\cite{poves1987,warburton1990} that were recently computed with ab initio methods~\cite{miyagi2020,novario2020,stroberg2021,Frosini:2021sxj,frosini2022,sun2024}. This is a significant step forward in mass number for ab initio computations of deformed nuclei.

{\it Methods.$-$}
We start from the intrinsic Hamiltonian
\begin{equation}
H=T-T_{\rm CoM}+V_{\rm NN}+V_{\rm 3N},
\end{equation}
where $T$ and $T_{\rm CoM}$ are the total kinetic energy and that of its center of mass, respectively. For the two-nucleon (NN) interaction $V_{\rm NN}$ and three-nucleon (3N) interaction $V_{\rm 3N}$ we use the chiral interaction 1.8/2.0(EM) \cite{hebeler2011}, which yields accurate ground-state energies and spectra of light, medium and heavy mass nuclei \cite{hagen2016b,simonis2017,morris2018,gysbers2019,stroberg2021,hu2022b,hebeler2023}. The NN interaction is calculated in the harmonic-oscillator basis with spacing $\hbar\omega$ and single-particle energies up to $N_{\rm max}\hbar\omega$; the 3N interaction is truncated to excitation energies of three nucleons up to $E_{\rm 3max}=28\hbar \omega$. The 3N forces were generated with the {\tt NuHamil} code~\cite{miyagi2023}. To overcome the computational challenge regarding the large number of 3N matrix elements, we used the normal-ordered two-body approximation \cite{hagen2007a,roth2012,djarv2021}, modified for deformed nuclei as follows~\cite{Frosini:2021tuj}. First, we performed a spherical Hartree-Fock calculation with the three-body force using a fractional filling of the orbit(s) at the Fermi surface. The three-body force is then normal ordered with respect to the Hartree-Fock vacuum and truncated at the two-body level. In a final step the normal-ordered Hamiltonian is transformed back to the harmonic oscillator basis. The ensuing Hartree-Fock and coupled-cluster calculations are based on this normal-ordered two-body interaction. They are described below.

We performed axially-symmetric Hartree-Fock computations using the Hamiltonian $H'=H-\lambda Q_{20}$ where $Q_{20}$ introduces the quadrupole deformation and $\lambda$ is a Lagrange multiplier. We followed Ref.~\cite{staszscak2010} to determine $\lambda$ for a given mass quadrupole deformation $\langle Q_{20}\rangle$. This allowed us to map out a potential energy surface (which is the Hartree-Fock energy of the Hamiltonian $H$ at a given deformation $\langle Q_{20}\rangle$). Each local minimum of the energy surface represents a distinct deformed configuration of the nucleus, and the corresponding Slater determinant is the reference state used for coupled-cluster computations. We need to account for short-range (dynamical) and long-range (static) correlations~\cite{sun2024}. We included the former via coupled-cluster with singles and doubles (CCSD) computations~\cite{shavittbartlett2009} and the latter via symmetry restoration~\cite{hagen2022,sun2024}. This last step allowed us to compute rotational bands. 
We computed \BE2 using the method developed in Ref.~\cite{sun2024}.

\begin{figure}
\setlength{\abovecaptionskip}{0pt}
\setlength{\belowcaptionskip}{0pt}
\includegraphics[scale=0.42]{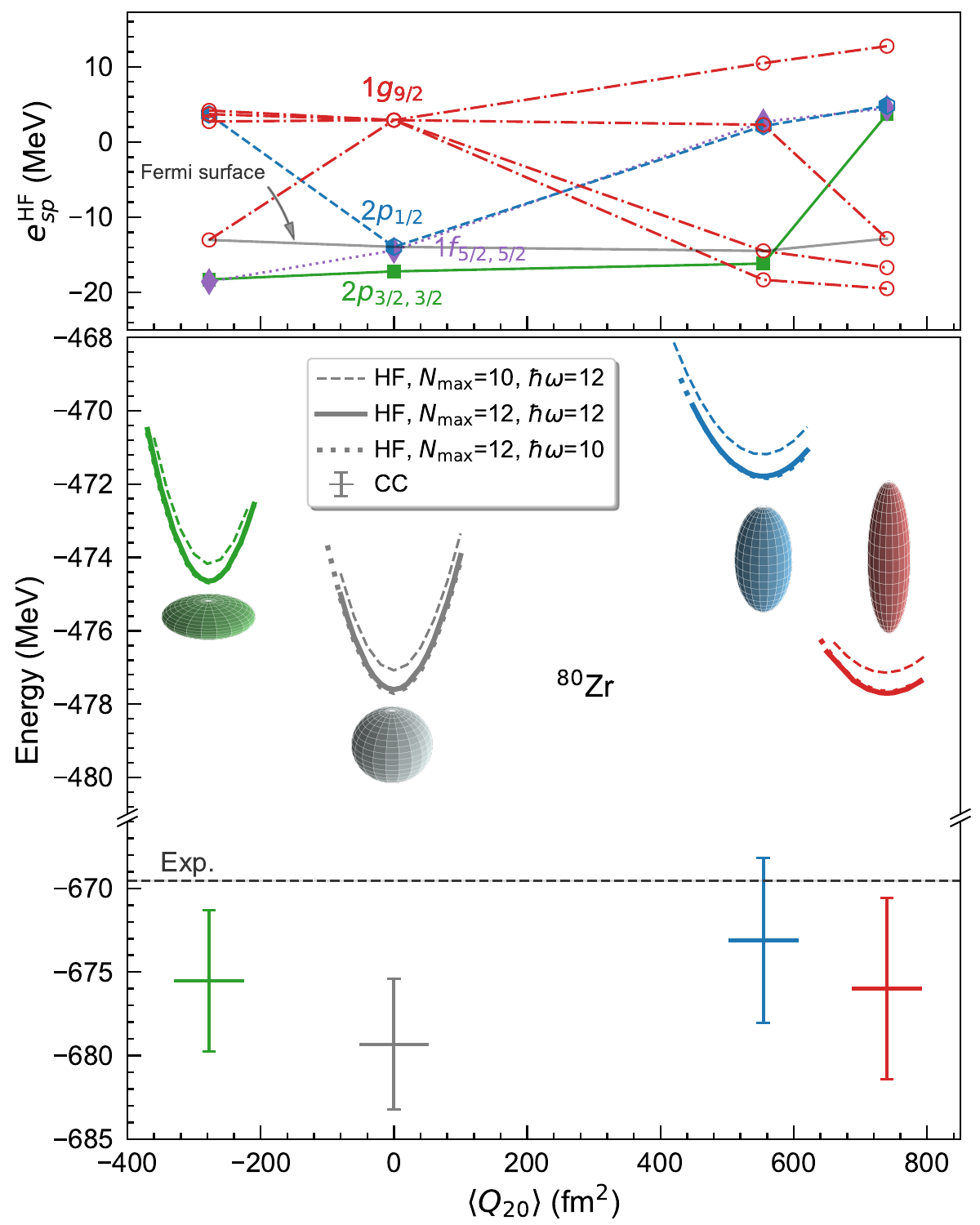}
\caption{\label{Zr80_EHF_Q20} Quadrupole constrained Hartree-Fock energies for different shapes in $^{80}$Zr (green, gray, blue, and red lines show oblate, spherical, prolate, and larger prolate configurations, respectively, for model-space sizes as indicated). Coupled-cluster results that include estimated energy contributions from triples excitations and angular momentum projection are shown in the lower part for the corresponding Hartree-Fock minima. Uncertainty estimates reflect model-space truncations. The horizontal gray dashed line shows the recent high-precision mass measurement of $^{80}$Zr~\cite{hamaker2021}.
The upper panel illustrates the Hartree-Fock single particle energy ($e_{sp}^{\rm HF}$) around the Fermi surface as a Nilsson diagram, with lines connecting the energies at the minima to guide the eye.}
\end{figure}

{\it Results.$-$}
The middle part of Fig.~\ref{Zr80_EHF_Q20} shows the unprojected Hartree-Fock energies of $^{80}$Zr in the vicinities of four minima. The oblate, spherical, prolate, and larger prolate minumum have deformation parameters $\beta_2 \approx$ $-$0.17, 0.0, 0.33 and 0.46, respectively. Here we obtained $\beta_2$ from the mass quadrupole moment $\langle Q_{20}\rangle$ via $\beta_2 \equiv \sqrt{5 \pi}\left\langle Q_{20}\right\rangle/(3 A R_0^2)$ with $R_0=1.2A^{1/3}$~fm~\cite{pritychenko2016}. It is interesting that the number and shapes of local Hartree-Fock minima computed with chiral interactions are consistent with the mean-field calculations~\cite{rodriguez2011}.

The lower part of Fig.~\ref{Zr80_EHF_Q20} shows coupled cluster results obtained at each minimum that include estimated energy contributions from triples excitations and angular momentum projection. Uncertainties come from model-space truncations. The energy from triples excitations is accurately estimated as 10\% of the CCSD correlation energy~\cite{sun2024}, the contribution from angular momentum projection was obtained from a projected CCSD calculation in a smaller model-space $N_{\rm max} = 6$, and the model-space uncertainties are estimated as the difference between the $N_{\rm max} = 10$ and $N_{\rm max} = 12$ of the unprojected CCSD results. The coupled-cluster calculations show that the spherical configuration is lowest in energy and slightly overbinds $^{80}$Zr when compared to data~\cite{hamaker2021}. However, within uncertainties it is not possible to unambiguously identify the ground-state shape between oblate, spherical, prolate, and larger prolate. Details are documented in Fig.~\ref{Energy_CC_HF} of the Supplemental Material.   

The upper part of Fig.~\ref{Zr80_EHF_Q20} shows how the different Hartree-Fock minima form a Nilsson diagram. The spherical state is a result of $N=Z=40$ harmonic-oscillator shell closure. The oblate configuration is obtained mainly from the inversion of \{2$p_{1/2}$ $\leftrightarrow$ 1$g_{9/2}$\} for both protons and neutrons at the Fermi surface. In the Nilsson diagram, the prolate states are from the level crossing, i.e.,  \{2$p_{1/2}$$\leftrightarrow$$1g_{9/2}$, 1$f_{5/2}$$\leftrightarrow$1$g_{9/2}$\} and  \{2$p_{1/2}$$\leftrightarrow$1$g_{9/2}$, 1$f_{5/2}$$\leftrightarrow$1$g_{9/2}$, 2$p_{3/2}$$\leftrightarrow$1$g_{9/2}$\} for the first and second prolate state, respectively. We also found an even more deformed ``super-prolate'' shape at $Q_{20} \approx 940$ fm$^2$ ($\beta_2 \approx$ 0.58) with \{2$p_{1/2}$$\leftrightarrow$1$g_{9/2}$, 1$f_{5/2}$$\leftrightarrow$1$g_{9/2}$, 2$p_{3/2}$$\leftrightarrow$2$d_{5/2}$\}. However, this deformation gave a too large \BE2 and a too compressed rotational band compared to the data, and we will omit it in this work.

We performed similar calculations of the nuclei $^{76}$Kr, $^{76,78}$Sr, $^{78}$Zr and $^{84}$Mo. Here again, the coexistence of various shapes results from the competition between the $N=Z=40$ harmonic oscillator shell closure and the intruding 1$g_{9/2}$ and 2$d_{5/2}$ orbitals. The convergence of the energies of different shapes is documented in Fig.~\ref{Energy_CC_HF} of the Supplemental Material. We find that method uncertainties make it difficult to unambiguously identify the lowest energy, i.e. the shape of the ground state. This makes it interesting to see how the spectra and values of \BE2 can be used to distinguish between the different shapes. 

\begin{figure*}
\setlength{\abovecaptionskip}{0pt}
\setlength{\belowcaptionskip}{0pt}
\begin{subfigure}{}
\includegraphics[scale=0.36]{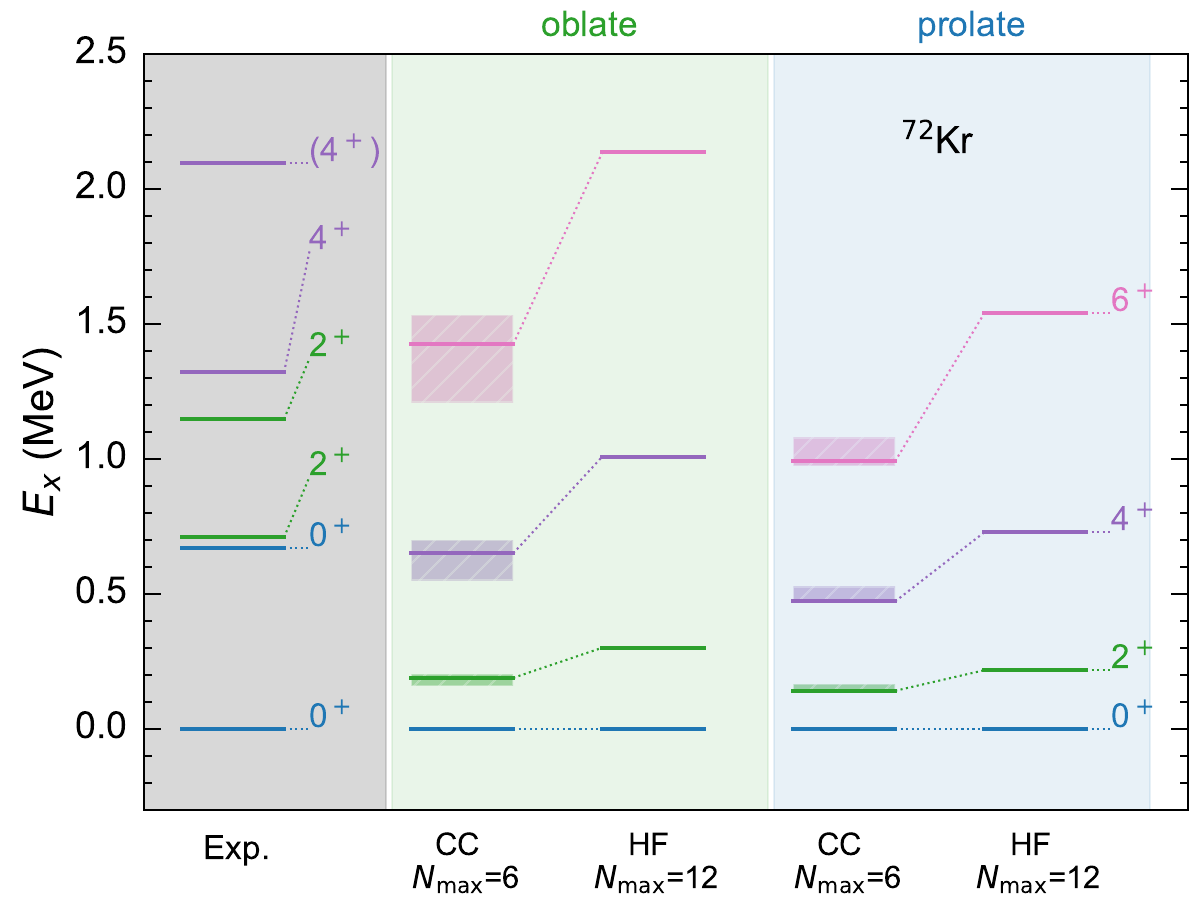}
\end{subfigure}
\begin{subfigure}{}
\includegraphics[scale=0.36]{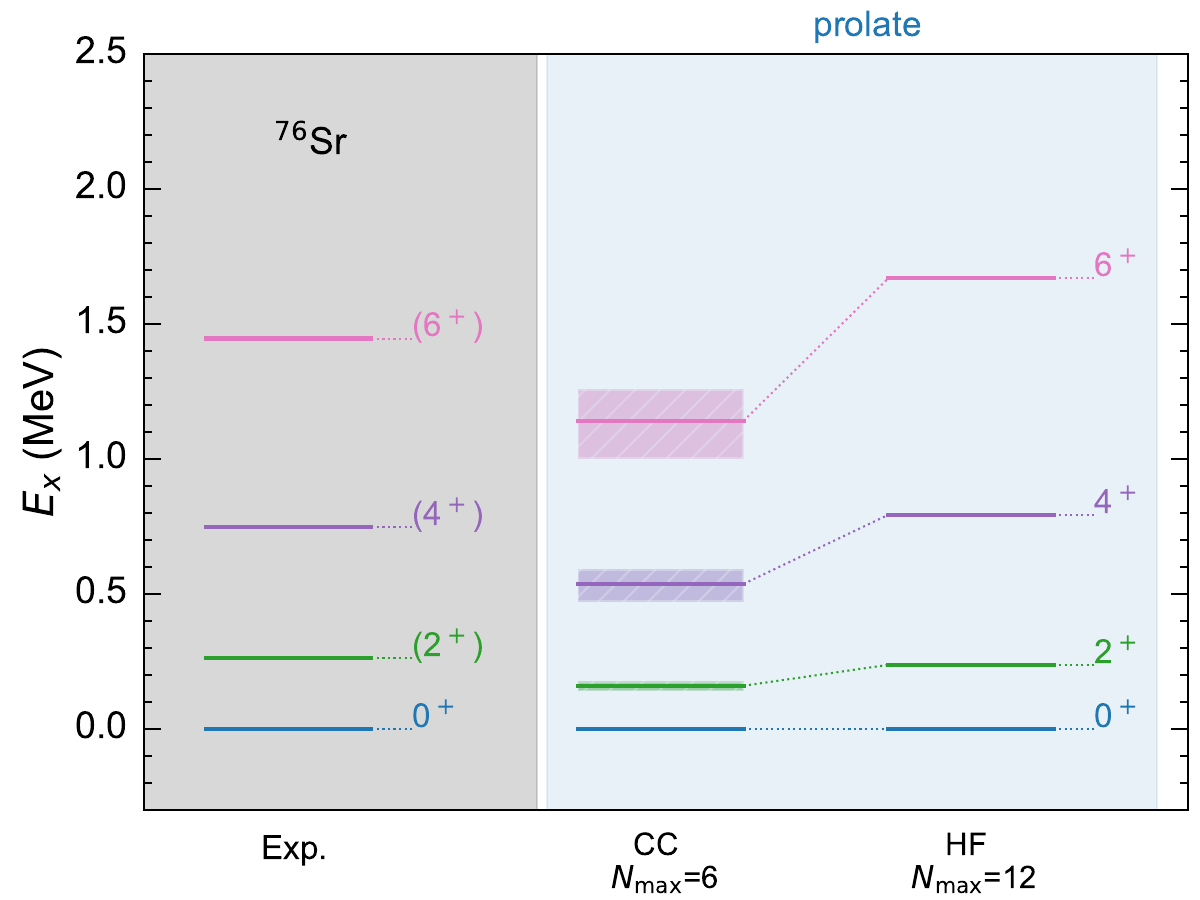}
\end{subfigure}
\\
\begin{subfigure}{}
\includegraphics[scale=0.36]{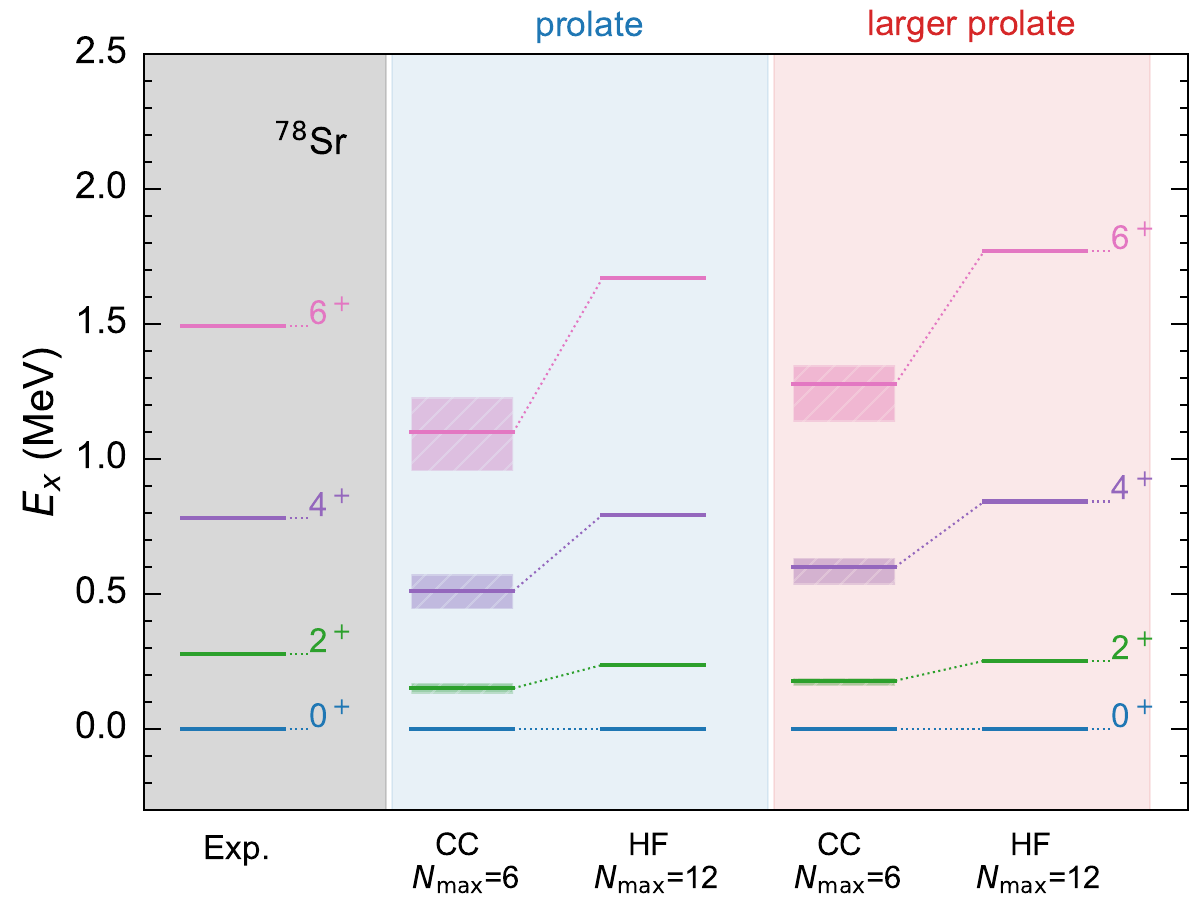}
\end{subfigure}
\begin{subfigure}{}
\includegraphics[scale=0.36]{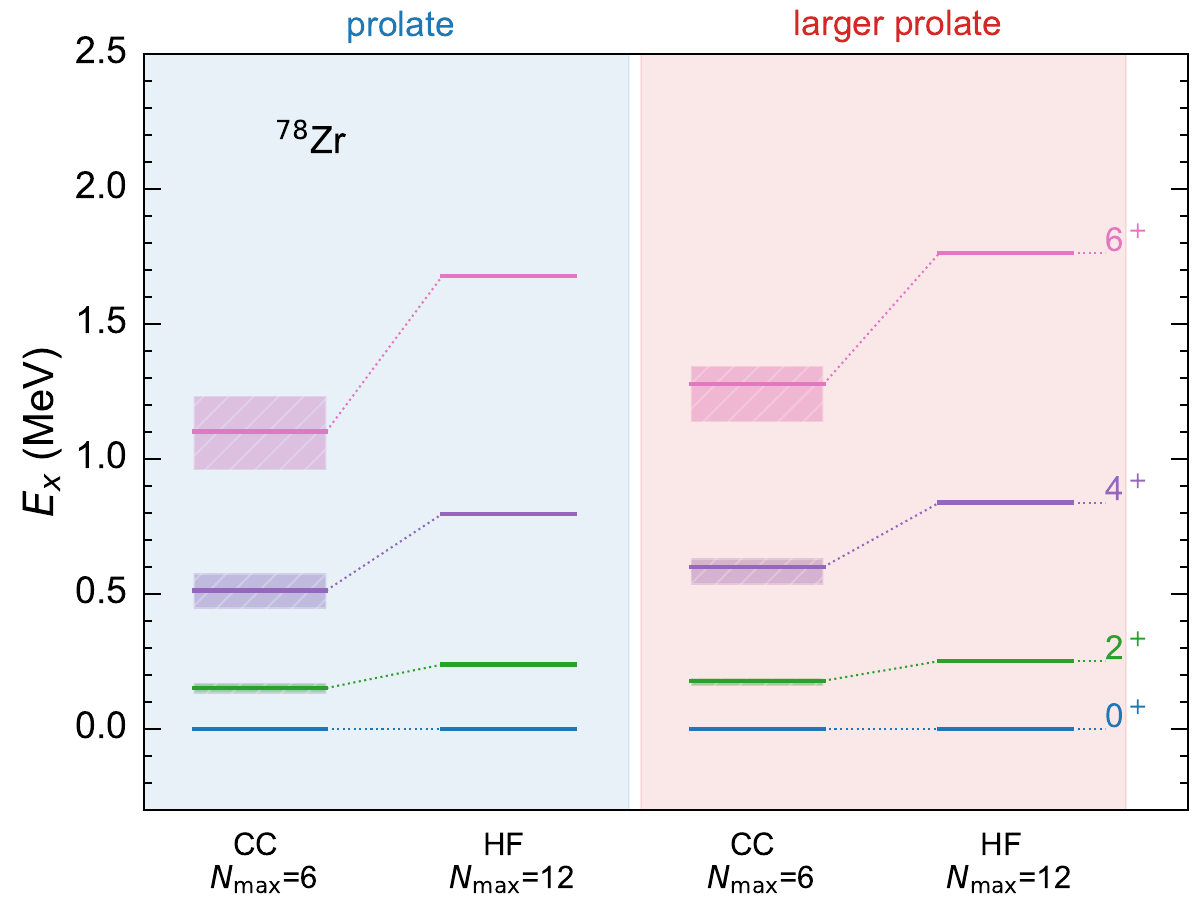}
\end{subfigure}
\\
\begin{subfigure}{}
\includegraphics[scale=0.36]{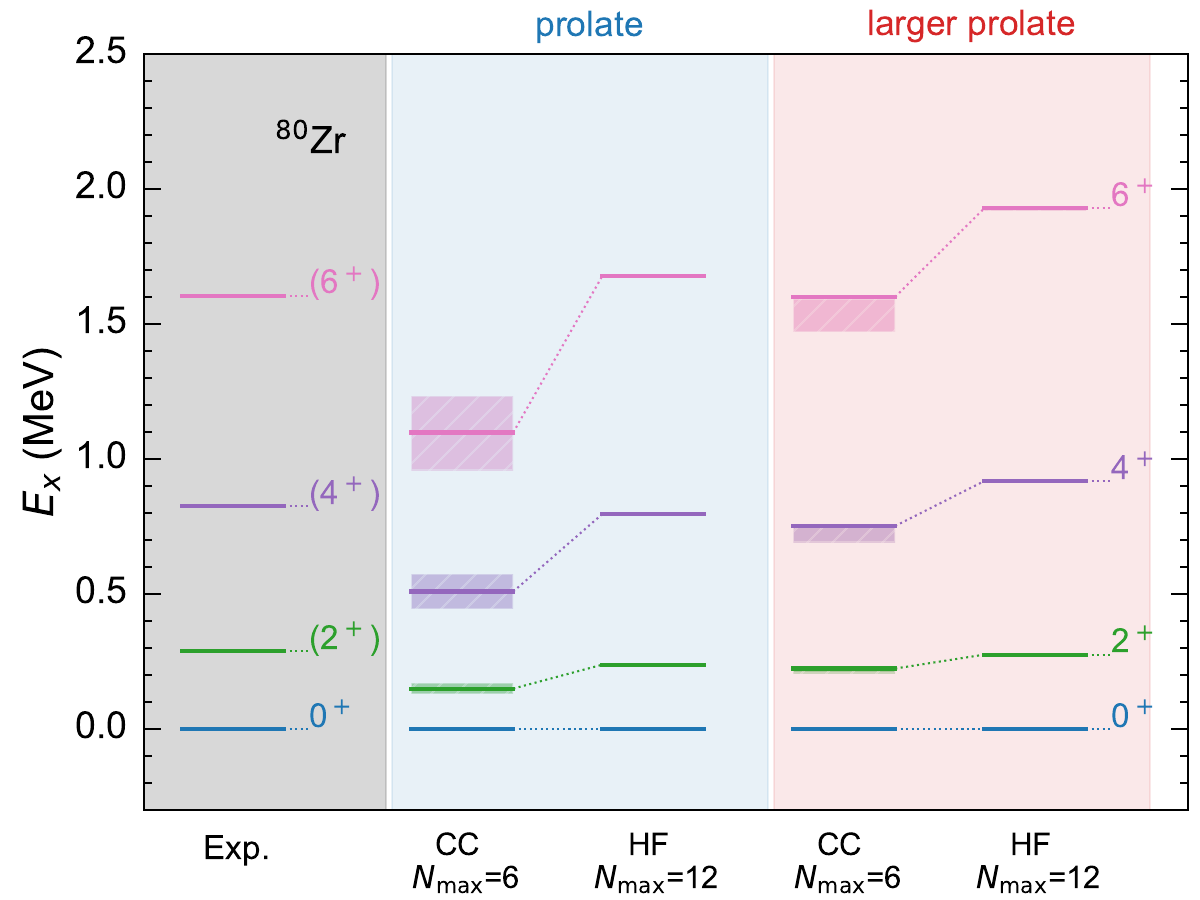}
\end{subfigure}
\begin{subfigure}{}
\includegraphics[scale=0.36]{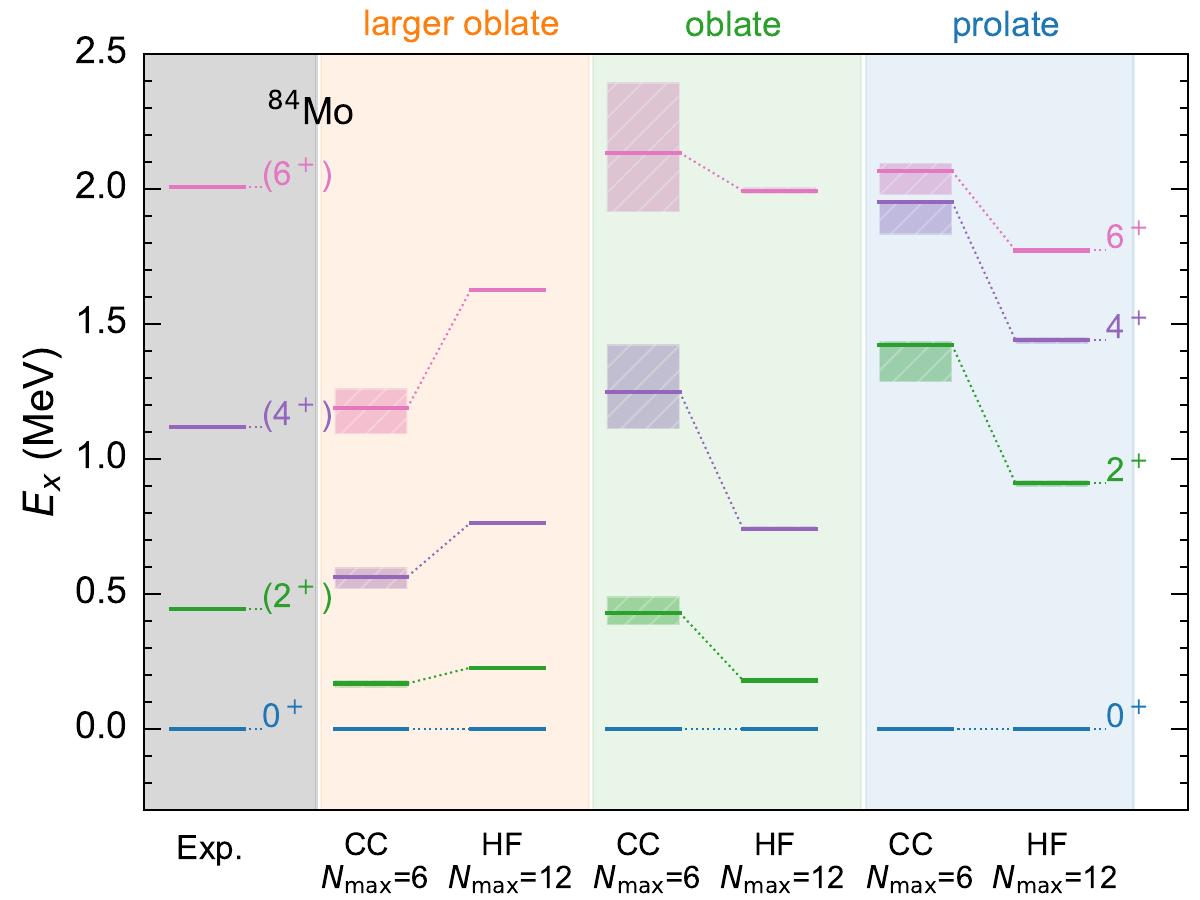}
\end{subfigure}
\caption{\label{Spectra_all} Excitation energies of $^{72}$Kr, $^{76,78}$Sr, $^{78,80}$Zr and $^{84}$Mo from the projected coupled cluster (CC) and projected Hartree Fock (HF) calculations for model spaces of size $N_{\rm max}$. The uncertainty is estimated based on varying the harmonic oscillator frequency $\hbar\omega$ from 10 to 16~MeV. Experimental data are from \cite{ensdf}.}
\end{figure*}

Figure~\ref{Spectra_all} shows the calculated rotational bands in $^{72}$Kr, $^{76,78}$Sr, $^{78,80}$Zr and $^{84}$Mo and compares them to data. The angular-momentum projection in coupled cluster theory is expensive and restricted to $N_{\rm max}=6$ for spectra and $N_{\rm max}=8$ for the \BE2's. Spectra from angular-momentum projections of axially symmetric Hartree-Fock states are well converged in such spaces, see Fig.~\ref{Spectra_Zr80} of the Supplemental Material. As  collective rotational phenomena are mainly related to long-range physics this suggests that the model spaces are large enough for the computation of rotational bands. In $^{80}$Zr, the spectrum agrees with that found for the larger prolate minimum. 

For $^{72}$Kr, the computed spectrum suggests that the ground state and its rotational band are oblate deformed. In reality, the situation is more complicated because oblate and prolate bands mix in this nucleus~\cite{bouchez2003,gade2005,iwasaki2014,wimmer2020}. 
We probed the mixing of different reference states in a generator-coordinate method. On the mean-field level, where we can assess such mixing, we see no evidence for it. In effective theories of nuclear deformation, mixing between two bands with identical $K$ quantum numbers is a higher-order effect~\cite{papenbrock2020}.  Thus, one needs pairs of levels with identical spins in two bands to be close in energy. In our case, the pairs of oblate and prolate $0^+$ states and $2^+$ states are about 1~MeV and 0.8~MeV apart, respectively. This is probably too large a separation. From this perspective, the mixing observed in $^{72}$Kr~\cite{wimmer2020} has some accidental character. 

In our calculations, the mirror nuclei $^{78}$Sr and $^{78}$Zr both exhibit two deformed Hartree-Fock minima and similar rotational bands. 
For $^{78}$Zr ($^{78}$Sr), \textcite{delaroche2010} found $2^+$ and $4^+$ states at 0.27 and 0.74~MeV (0.30 and  0.79~MeV), respectively. Although our projected Hartree-Fock results are close to these and the data in $^{78}$Sr, the coupled-cluster spectra are too compressed. 
For $^{84}$Mo, \textcite{delaroche2010} found $2^+$ and $4^+$ states at 0.54 and  1.20~MeV, respectively, and this is close to data. Our coupled-cluster results for the oblate deformation agree with these results.

With the exception of molybdenium, our projected Hartree-Fock results appear easier to match with data than the projected coupled-cluster ones. We can only speculate about possible shortcomings in our computations.  First, we are limited to axial symmetry. Any static triaxial deformation or $\gamma$-softness cannot be captured with our present limitation. Second, in the angular-momentum projection we approximate a product of the rotation operator and the exponential cluster-excitation operator as a new exponential including up to two-particle--two-hole excitations~\cite{qiu2017,hagen2022,sun2024}. The inclusion of three-particle--three-hole terms is beyond our computational abilities because those amplitudes lack axial symmetry and require too large memory demands. Although this approximation was accurate in light nuclei up to mass number 34~\cite{sun2024}, it is otherwise hard to assess its precision.     

\begin{figure}
\setlength{\abovecaptionskip}{0pt}
\setlength{\belowcaptionskip}{0pt}
\includegraphics[scale=0.38]{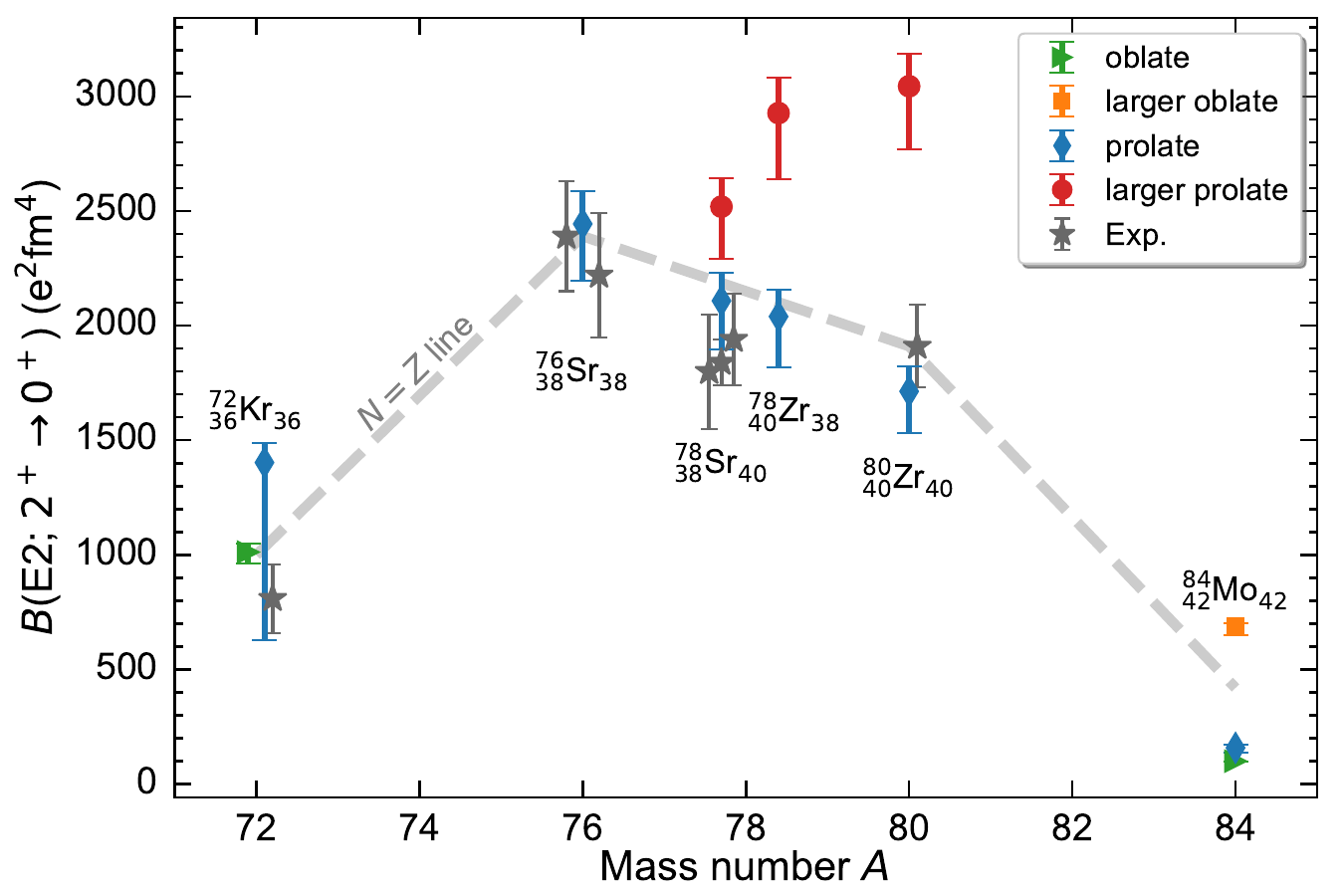}
\caption{\label{BE2}$B({\rm E}2;2^+_1 \rightarrow 0^+_1)$ values for the $N$=$Z$ nuclei from $^{72}$Kr to $^{84}$Mo. The values for mirror nuclei $^{78}$Sr and $^{78}$Zr are also illustrated. The uncertainty is estimated based on the basis parameter $\hbar\omega$, ranging from 10 to 14 MeV. Experimental data are from Refs.~\cite{lister1982,iwasaki2014,lemasson2012,llewellyn2020}.  
}
\end{figure}

We turn to the reduced transition strengths \BE2 as another measure of nuclear deformation. Figure~\ref{BE2} shows the \BE2 values for nuclei computed in this work and compares them to the data. The coupled-cluster results of $^{72}$Kr and $^{76}$Sr agree  with the data. The \BE2 of $^{78}$Sr and $^{80}$Zr suggest that the prolate state is the ground state. 
The larger-prolate configuration reproduces a relatively correct spectrum but too large \BE2. 
The \BE2 values of the mirror nuclei pair $^{78}$Sr and $^{78}$Zr, shown in Fig.~\ref{BE2}, agree within uncertainties for both deformations.

\begin{figure}
\setlength{\abovecaptionskip}{0pt}
\setlength{\belowcaptionskip}{0pt}
\includegraphics[scale=0.38]{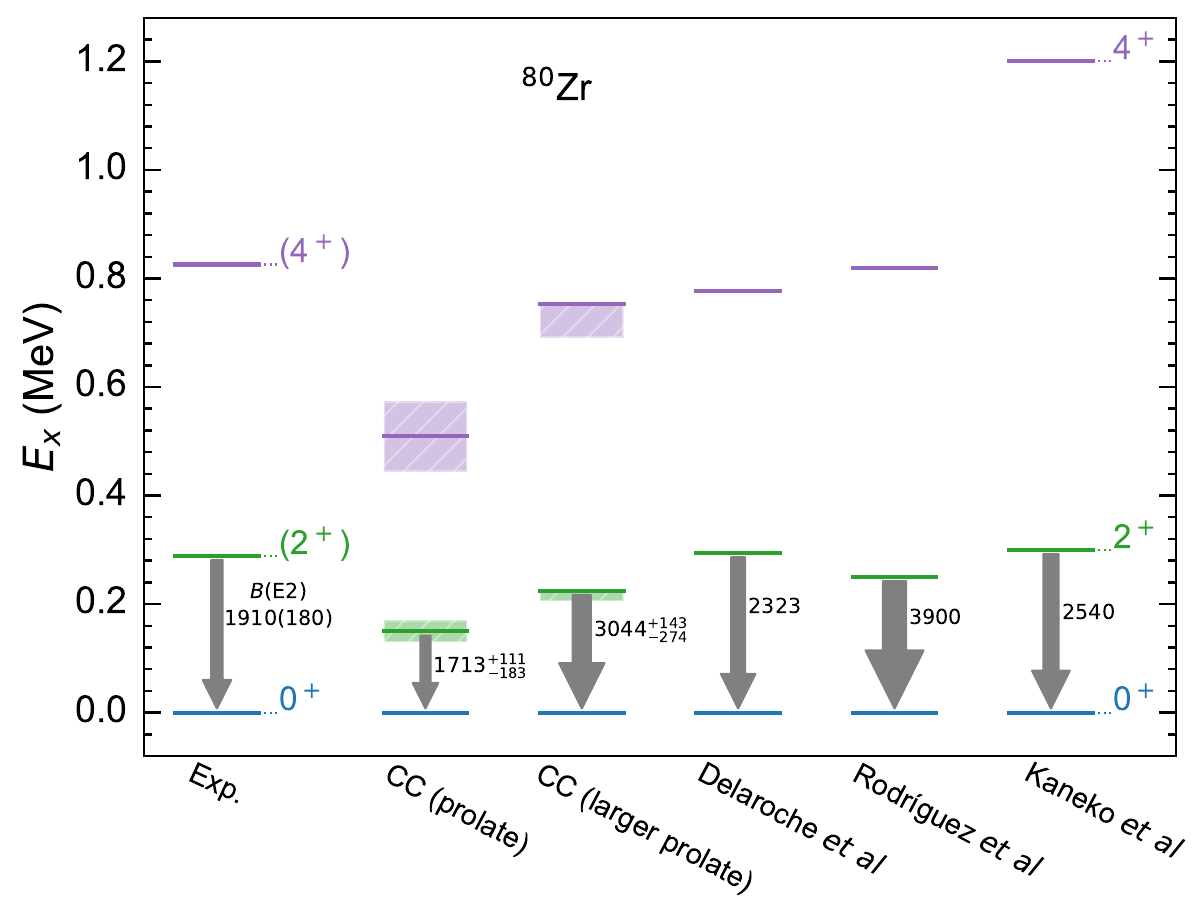}
\caption{\label{Zr80-compare} Spectrum of $^{80}$Zr (Exp.) compared to the coupled-cluster (CC) results of this work, and to the results from \textcite{delaroche2010}, \textcite{rodriguez2011}, and \textcite{kaneko2021}. $B({\rm E}2;2^+_1 \rightarrow 0^+_1)$ values fare also indicated.}
\end{figure}

Figure~\ref{Zr80-compare} compares the data for $^{80}$Zr with the results from this work and other models. Clearly, this nucleus is challenging, and the results by \textcite{delaroche2010} are overall closest to data. 
We continue the comparison of our ab initio approach with other models and compiled data in Table~\ref{tab1}. Here, we singled out the beyond-mean-field computations by \textcite{delaroche2010} in column~4 and showed results from other computations in the last column. The main point is here that our calculations reach an accuracy that is comparable to other nuclear models.

\setlength{\tabcolsep}{6pt}
\renewcommand{\arraystretch}{1.5} 
\begin{table}[!htb]
 \centering
 \begin{tabular}{|l r r r r|}
 \hline
\hline
Nucleus &~Exp. &~This work & Ref.~\cite{delaroche2010} & Other\\
\hline
\multirow{2}{*}{$^{80}$Zr} & \multirow{2}{*}{1910(180)$^a$}& $1713^{+111}_{-183}$ & \multirow{2}{*}{2323} & $3900^b$\\
& & $3044^{+143}_{-274}$ & & 2540$^f$\\
\hline

\multirow{2}{*}{$^{78}$Zr} & \multirow{2}{*}{not known} & $2040^{+118}_{-220}$ & \multirow{2}{*}{2504} & \\
& & $2927^{+155}_{-288}$ & & \\
\hline

\multirow{2}{*}{$^{78}$Sr} & \multirow{2}{*}{1840(100)$^a$} & $2108^{+121}_{-211}$ & \multirow{2}{*}{1989} & 2291$^f$\\
& & $2519^{+125}_{-228}$ & & \\
\hline

$^{76}$Sr & 2390(240)$^a$ & $2444^{+145}_{-248}$ & 2350 & 2175$^f$\\
\hline

\multirow{2}{*}{$^{72}$Kr} & 810(150)$^c$& $1012^{+36}_{-50}$& \multirow{2}{*}{819} & 763$^d$\\
& 999(129)$^e$ & $1403^{+84}_{-775}$ & & 1097$^f$ \\
 \hline
 \hline
\end{tabular}
\caption{$B({\rm E}2; 2^+\to0^+)$ values (in units of $e^2{\rm fm}^4$) from experiment, this work, \textcite{delaroche2010}, and other references for various nuclei. $a =$~Ref.~\cite{llewellyn2020}; $b =$~Ref.~\cite{rodriguez2011}; $c =$~Ref.~\cite{iwasaki2014}; $d =$~Ref.~\cite{bender2006}; $e =$~Ref.~\cite{gade2005}; $f =$~Ref.~\cite{kaneko2021}, with effective charges adjusted to \BE2. The values reported under ``This work'' are those presented in Fig.~\ref{BE2} and reflect the shapes and deformation as indicated there.}
\label{tab1}
\end{table}

\begin{figure}
\setlength{\abovecaptionskip}{0pt}
\setlength{\belowcaptionskip}{0pt}
\includegraphics[scale=0.42]{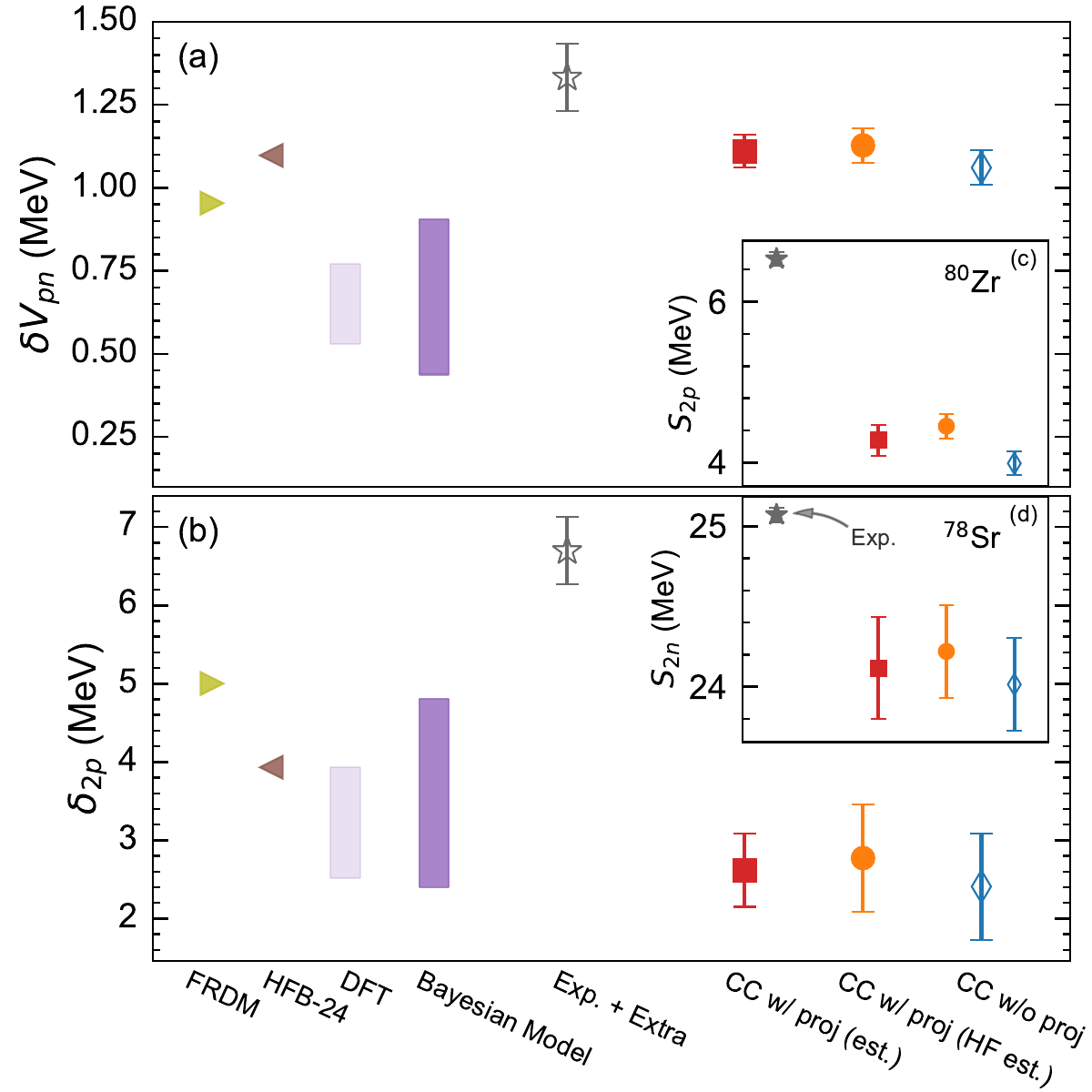} \caption{\label{Mass_anomaly} Double mass difference $\delta V_{pn}$ (top), two-proton shell gap $\delta_{2p}$ (bottom), and two-proton/neutron separation energy $S_{2p,2n}$ (insets), from FRDM~\cite{moller2016}, HFB-24~\cite{goriely2013}, our coupled-cluster (CC) computations, and the DFT, Bayesian Model, and experiment-plus-extrapolation (Exp.+Extra) results from Ref.~\cite{hamaker2021}. The coupled-cluster uncertainties are estimated based on the harmonic oscillator frequency $\hbar\omega$, ranging from 10 to 16 MeV. 
The experimental data are from Ref.~\cite{Wang_2021}.  
}
\end{figure}

We also want to get a sense on how our results depend on the chosen interaction. 
We remark that, although 1.8/2.0(EM) interaction underestimates nuclear radii \cite{hagen2015,hagen2016b,simonis2017}, it does not significantly underpredict \BE2 values~\cite{sun2024}. For example, our \BE2 examination of the larger prolate shape ($\langle Q_{20}\rangle = 737$~fm$^2$) in $^{80}$Zr, calculated in a model space of $N_{\rm max}$=8 and $\hbar\omega$=12 MeV, yields a value of 3044~$e^2$fm$^4$. The $\Delta$NNLO$_{\rm GO}$~\cite{jiang2020} interaction, which more accurately reproduces  charge radii, yields a \BE2 value of 3449~$e^2$fm$^4$ at the larger prolate minimum ($\langle Q_{20}\rangle = 859$~fm$^2$), see Fig.~\ref{interactions} in Supplemental Material. The difference between the \BE2 and $\langle Q_{20}\rangle$ values for the two different interactions is consistent with the difference in charge radii~\cite{jiang2020,groote2020}. We note the rotational band is more compressed for the $\Delta$NNLO$_{\rm GO}$ interaction, consistent with expectations from $\langle Q_{20}\rangle$.

Finally, we revisited the mass anomaly near $N$=$Z$=40 region reported by \textcite{hamaker2021}.  That work showed that theoretical approaches  struggle to reproduce the four-point mass difference $\delta V_{pn}$ = $\frac{1}{4}[B(N,Z)-B(N-2,Z)-B(N,Z-2)+B(N-2,Z-2)]$ and the two-proton shell gap $\delta_{2p}$=$2B(N,Z)-B(N,Z+2)-B(N,Z-2)$. Here $B(N,Z)$ is the binding energy of a nucleus with proton number $Z$ and neutron number $N$. To extract $\delta V_{pn}$ and $\delta_{2p}$ at $N$=$Z$=40, one needs the masses of $^{76}$Sr, $^{78}$Sr, $^{78,80}$Zr and $^{82}$Mo (which we also calculated, see Fig.~\ref{Energy_CC_HF} in Supplemental Material). As only $^{76}$Sr, $^{78}$Sr, $^{80}$Zr have been measured, \textcite{hamaker2021} used extrapolated masses for $^{78}$Zr and $^{82}$Mo. 

Figure~\ref{Mass_anomaly} shows our extracted $\delta V_{pn}$ and $\delta_{2p}$. 
Our results are consistent with previous calculations from Ref.~\cite{hamaker2021} using the finite-range droplet mass model (FRDM), the density functional theory (DFT), as well as Bayesian analysis based on these models. All theoretical results of $\delta V_{pn}$ and $\delta_{2p}$ deviate from those derived from data and extrapolated masses. Additional measurements for $^{78}$Sr and $^{82}$Mo are required to confirm this anomaly.

Considering the separation energies, data are available for the two-proton separation energy $S_{2p}$=$B(N,Z)-B(N,Z-2)$ in $^{80}$Zr, and the two-neutron separation energy $S_{2n}$=$B(N,Z)-B(N-2,Z)$ in $^{78}$Sr. The inset in Fig.~\ref{Mass_anomaly} shows our CCSD results for $S_{2p}$ and $S_{2n}$, and as can be seen they are smaller then data. This discrepancy might be due to missing correlations in the CCSD approximation, the employed interaction, and the normal-ordered two-body approximation of the 3N interaction.

{\it Discussion.$-$}
On the one hand, ab initio computations can now describe shape coexistence in nuclei with mass numbers of about 80. On the other hand, they lack the precision to unambiguously determine nuclear ground-state shapes. This challenge exists because the 1\% uncertainties in the total energies for different shapes are larger than the small differences between them. As is documented in the literature~\cite{stoitsov2013,schunck2015,marevic2022}, nuclear energy functionals could be facing similar challenges (though model-space uncertainties were not reported in Refs.~\cite{delaroche2010,rodriguez2011}). \textcite{schunck2015}, for instance,  showed that the model-space dependence of the ground-state energy increases from less than 1~MeV in $^{40}$Ca to 7~MeV in $^{240}$Pu. \textcite{marevic2022} showed that it might be difficult to decide the groundstate shape of $^{50}$Cr because different bases yield different results. There is also good news, and we mention two points. First, the \BE2 values from symmetry-projected computations are improved compared to the valence-space IMSRG where these are typically too small~\cite{parzuchowski2017,miyagi2020,stroberg2022}. Second, the ab initio results are of similar quality as from comparable mean-field calculations. 

{\it Summary.$-$}
We investigated the low-lying collective states and \BE2 values in the heavy $N$=$Z$ region using the ab-initio coupled cluster calculations based on axially-symmetric reference states followed by angular-momentum projection. While we found coexistence between various oblate and prolate shapes the calculations were not precise enough to unambiguously identify the shape of the ground states. In  particular, we are unable to simultaneously reproduce the rotational band and \BE2 for $^{78}$Sr and $^{80}$Zr. This discrepancy of theoretical calculations and data in these region remains a challenge to nuclear theories, and poses the need for further theoretical development. The computations presented here provide us with a useful step towards the description of deformed nuclei in heavy-ion collisions \cite{jia2023}, and for tests of fundamental symmetries \cite{ENGEL201321}.
\newline{}  

\begin{acknowledgments}
We thank Takayuki Miyagi for the {\tt NuHamil} code~\cite{miyagi2023} and Ragnar Stroberg for the {\tt imsrg++} code~\cite{Stro17imsrg++} used to generate matrix elements of the chiral three-body interaction. This work was supported by the U.S. Department of Energy (DOE), Office of
Science, under SciDAC-5 (NUCLEI collaboration), under grant DE-FG02-97ER41014, and by the Quantum Science Center, a National Quantum Information Science Research Center of the U.S. Department of Energy.
This research used resources from the Oak Ridge Leadership Computing Facility located at Oak Ridge National Laboratory, which is supported by the Office of Science of the U.S. Department of Energy under contract No. DE-AC05-00OR22725.
\end{acknowledgments}

\bibliography{Ref_Zr80,Ref_all,master}

\newpage
\clearpage
\section{\label{Supplementary}Supplemental Material}
\begin{figure}[h!]
\setlength{\abovecaptionskip}{0pt}
\setlength{\belowcaptionskip}{0pt}
\includegraphics[scale=0.32]{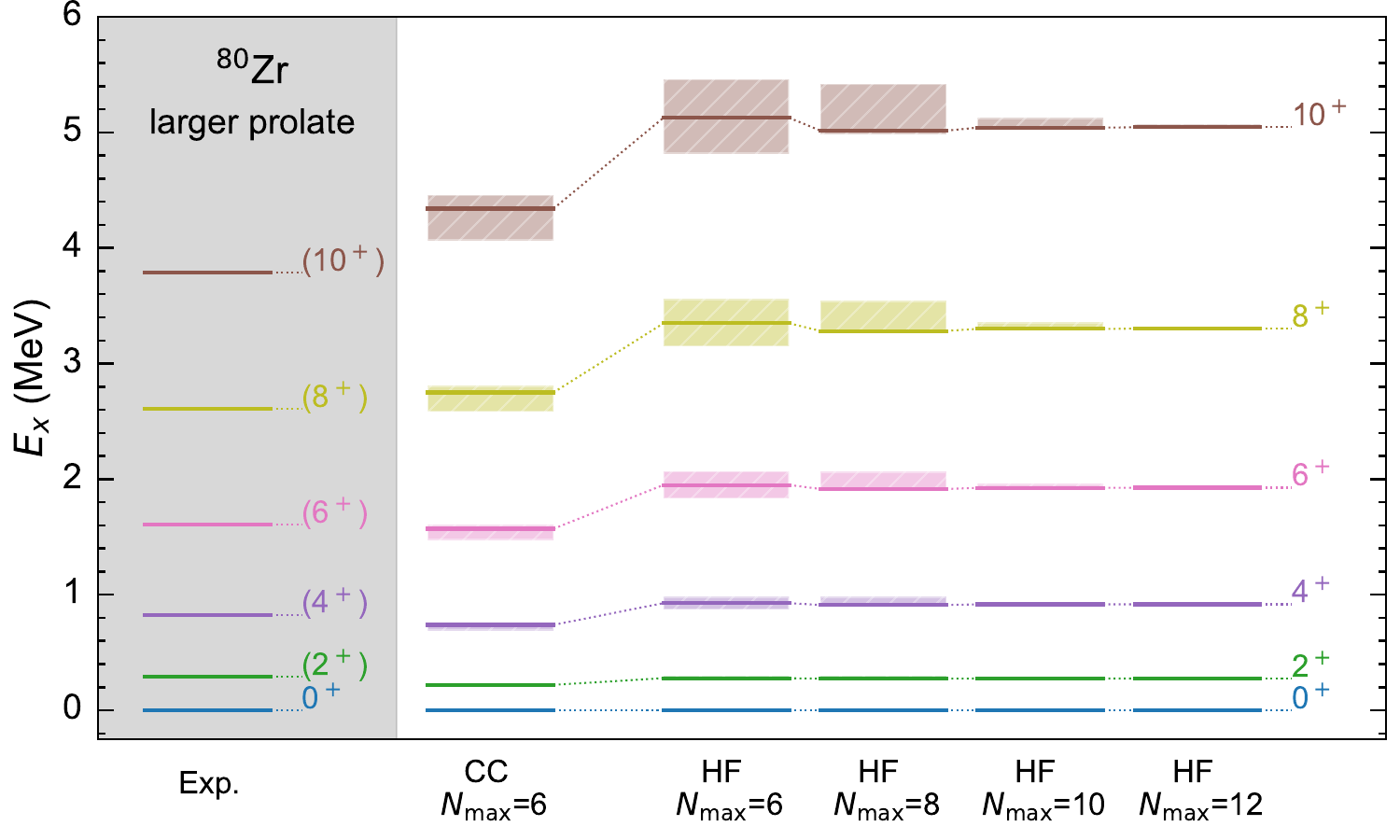}
\caption{\label{Spectra_Zr80} Rotational band in $^{80}$Zr from coupled cluster and Hartree-Fock calculations for different model spaces $N_{\rm max}$ as indicated, and compared to data.}
\end{figure}

Figure~\ref{Spectra_Zr80} shows that the rotational band in $^{80}$Zr is converged for $N_{\rm max} = 6$ for the projected Hartree-Fock results. The uncertainty bands reflect stem from varying the harmonic oscillator frequency $\hbar\omega$ from 10 to 16~MeV. Projectecd coupled cluster calculations are based on the $N_{\rm max} = 6$ model space. 

\begin{figure}[h!]
\setlength{\abovecaptionskip}{0pt}
\setlength{\belowcaptionskip}{0pt}
\includegraphics[scale=0.42]{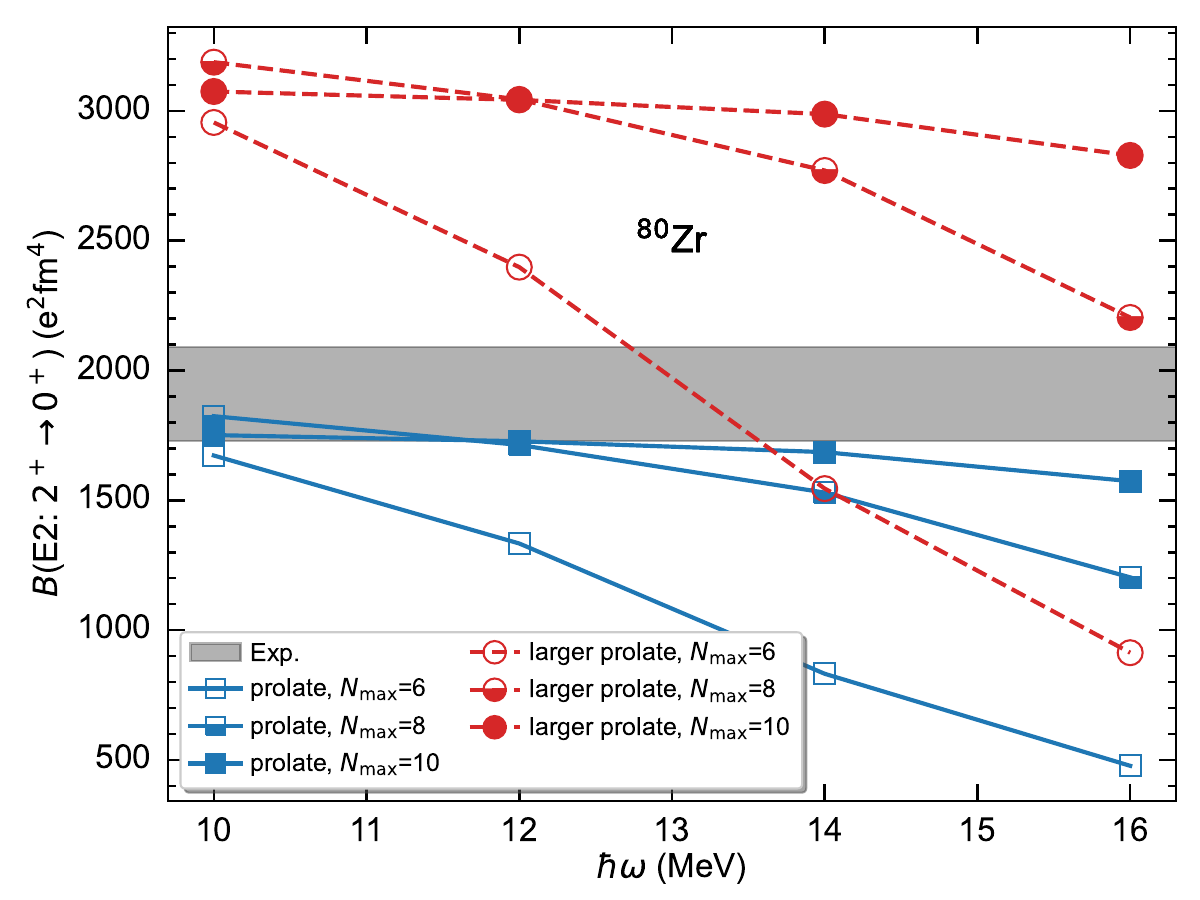}
\caption{\label{BE2_Zr80} The $B({\rm E}2;2^+_1 \rightarrow 0^+_1)$ in $^{80}$Zr for two prolate deformations and different model space sizes $N_{\rm max}$ as a function of the harmonic oscillator frequency $\hbar\omega$, and compared to data.}
\end{figure}

Figure~\ref{BE2_Zr80} shows the \BE2 in $^{80}$Zr for two prolate deformations and different model spaces $N_{\rm max}$ as a function of the harmonic oscillator frequency $\hbar\omega$.

\begin{figure}[h!]
\setlength{\abovecaptionskip}{0pt}
\setlength{\belowcaptionskip}{0pt}
\includegraphics[scale=0.42]{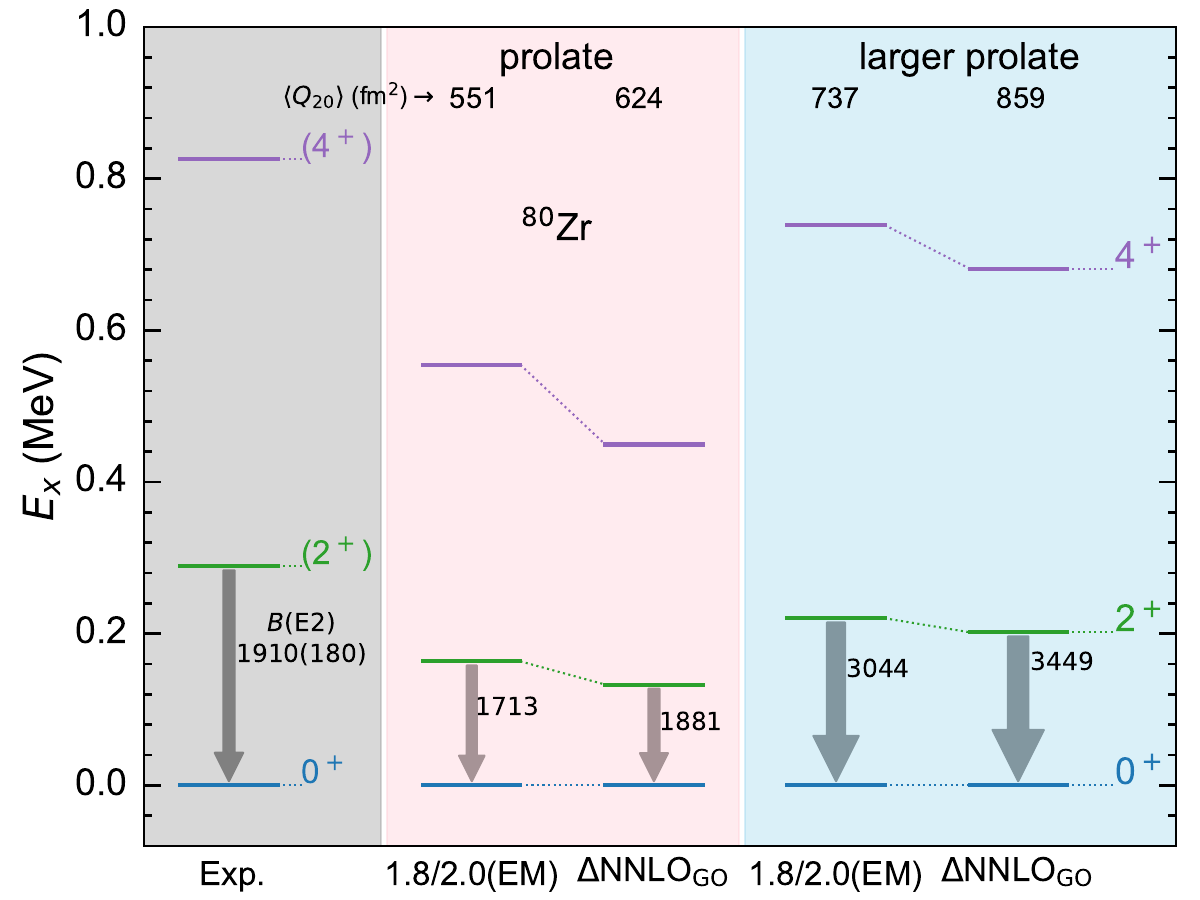}
\caption{\label{interactions} Coupled cluster results of spectra and $B({\rm E}2;2^+_1 \rightarrow 0^+_1)$ in $^{80}$Zr using the 1.8/2.0(EM) and $\Delta$NNLO$_{\rm GO}$ interactions, and compares them to data.}
\end{figure}

Figure~\ref{interactions} presents coupled cluster results for $^{80}$Zr using the 1.8/2.0(EM) and $\Delta$NNLO$_{\rm GO}$ interactions, and compares them to data. 

\begin{figure*}
\setlength{\abovecaptionskip}{0pt}
\setlength{\belowcaptionskip}{0pt}
\includegraphics[scale=0.90]{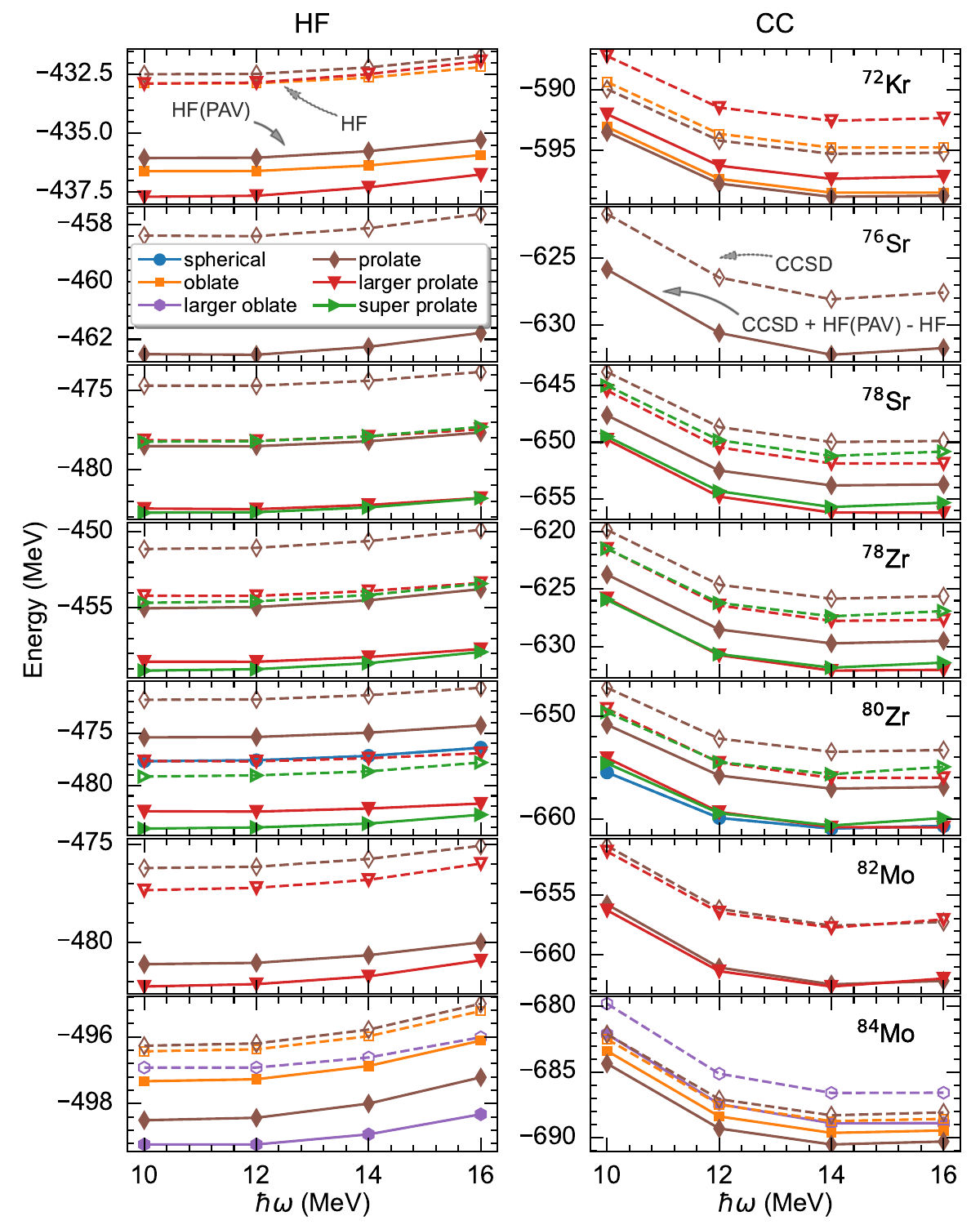}
\caption{\label{Energy_CC_HF} Energies of $^{72}$Kr, $^{76,78}$Sr, $^{78,80}$Zr and $^{82,84}$Mo (from top to bottom) from  coupled cluster (left) and Hartree Fock (right) calculations as a function of the harmonic-oscillator frequency $\hbar\omega$. Different colors indicate different deformations. Dashed lines are from symmetry-breaking calculations while full lines are from symmetry projection [labeled HF(PAV)]. For the projected coupled cluster results we simply added the energy differences between HF(PAV) and Hartree-Fock to the unprojected coupled cluster results. The model space has $N_{\rm max} =12$.}
\end{figure*}

Figure~\ref{Energy_CC_HF} shows the projected and unprojected Hartree-Fock and coupled-cluster ground-state energies in $^{72}$Kr, $^{76,78}$Sr, $^{78,80}$Zr and $^{82,84}$Mo for different oblate and prolate deformations as a function of the harmonic oscillator frequency $\hbar\omega$. 

\newpage 

\end{document}